\documentclass[submission,copyright]{eptcs}
\providecommand{\event}{HCVS 2023} %

\usepackage[bottom]{footmisc}
\usepackage{iftex}
\usepackage{float}
\usepackage{graphicx}
\usepackage{amsfonts,amsmath,amssymb}
\usepackage{listings}
\usepackage[hang]{caption}
\usepackage{subcaption}
\usepackage[english]{babel}
\usepackage{multirow}
\usepackage{tikz}
\usepackage{orcidlink}

\usetikzlibrary{positioning, calc, shapes, shapes.misc, shapes.arrows, external, patterns, arrows.meta}

\usepackage{pgfplots}
\usepackage{pgfplotstable}
\pgfplotsset{compat=1.7}

\ifpdf
  \usepackage{underscore}         %
  \usepackage[T1]{fontenc}        %
\else
  \usepackage{breakurl}           %
\fi

\title{Bottoms Up for CHCs: Novel Transformation of Linear Constrained Horn Clauses to Software Verification}
\author{M\'ark Somorjai\orcidlink{0000-0001-7537-0469} \qquad Mih\'aly Dobos-Kov\'acs\orcidlink{0000-0002-0064-2965} \qquad Zs\'ofia \'Ad\'am\orcidlink{0000-0003-2354-1750} \\ Levente Bajczi\orcidlink{0000-0002-6551-5860} \qquad Andr\'as V\"or\"os\orcidlink{0000-0001-7617-3563}
\email{vori@mit.bme.hu} 
\institute{
Department of Measurement and Information Systems\\
Budapest University of Technology and Economics}
}
\def\titlerunning{Novel Transformation of Linear Constrained Horn Clauses to Software Verification}
\def\authorrunning{Márk Somorjai et al.}

\begin{document}
\maketitle

\newtheorem{example}{Example}
\newcommand{\exampleautorefname}{Example}
\newtheorem{definition}{Definition}
\newtheorem{theorem}{\tetel}
\renewcommand{\figureautorefname}{Figure}
\renewcommand{\tableautorefname}{Table}
\renewcommand{\sectionautorefname}{Section}
\renewcommand{\subsectionautorefname}{Subsection}

\begin{abstract}
Constrained Horn Clauses (CHCs) have conventionally been used as a low-level representation in formal verification. 
Most existing solvers use a diverse set of specialized techniques, including direct state space traversal or under-approximating abstraction, necessitating purpose-built complex algorithms.
Other solvers successfully simplified the verification workflow by translating the problem to inputs for other verification tasks, leveraging the strengths of existing algorithms.
One such approach transforms the CHC problem into a recursive program roughly emulating a \emph{top-down} solver for the deduction task; and verifying the reachability of a safety violation specified as a control location.
We propose an alternative \emph{bottom-up} approach for linear CHCs, and evaluate the two options in the open-source model checking framework \textsc{Theta} on both synthetic and industrial examples.
We find that there is a more than twofold increase in the number of solved tasks when the novel \emph{bottom-up} approach is used in the verification workflow, in contrast with the \emph{top-down} technique.
\end{abstract}
\section{Introduction}
Constraint Horn Clauses (CHCs) are widely used in the field of formal verification both as a means for an intermediate representation \cite{tricera, seahorn, rusthorn} and as a specification language \cite{chc-comp_report22}. Conventionally, CHCs allow the specification of \emph{deduction} problems using implication, allowing the formalization of rules that govern how atomic facts lead to more complex (\emph{deduced}) information. 

A CHC problem can be solved in many different ways. \textsc{Spacer} \cite{spacer} in \textsc{Z3} \cite{z3} uses a solver based on automatic under-approximating abstraction; \textsc{Eldarica} \cite{eldarica} uses a direct abstract state space traversal over the CHC formulae; and \textsc{Unihorn} \cite{chc-comp_report22} uses a translation to recursive Boogie \cite{boogie} code before applying a conventional software verification workflow to achieve a result. While the former approaches in \textsc{Spacer} and \textsc{Eldarica} work well as demonstrated by their performance in previous years' CHC-COMP \cite{chc-comp_report22}, a competition for CHC solvers, they require purpose-built solvers, thus incurring additional effort when developing new algorithms.  

In contrast, the approach utilized by \textsc{Unihorn} relies on existing algorithms, taking advantage of the tool being part of the \textsc{Ultimate} framework with proven and efficient algorithms for tackling software verification tasks \cite{automizer}. By complementing the framework with a new front-end for parsing and transforming CHC formulae, the same verification workflows can be applied to the CHC-based problems as well, enabling their efficient verification. 

The transformation step used by \textsc{Unihorn} creates Boogie code that roughly emulates a program capable of deducing the existence of facts necessary to reach some end goal (e.g., a safety violation). We refer to this approach as \emph{top-down} \cite{datalog_bottomup} or \emph{backward}.

In this paper, we introduce an alternative to the \emph{backward} method, which creates a program that emulates a \emph{bottom-up} solver \cite{datalog_bottomup} (i.e., starting from nondeterministic facts and trying to deduce a safety violation using the formulae). We implement this \emph{forward} transformation to another formal representation of programs, the Control Flow Automaton (CFA), alongside with a \emph{backward} transformation alternative, in \textsc{Theta} \cite{theta}. Our benchmarks show that using the proposed approach increased the number of successfully solved CHCs more than twofold on linear CHC verification tasks from the CHC-COMP benchmark suite \cite{chc-comp_report22}.

This paper is structured as follows. In \autoref{sec:background}, we introduce the necessary background concepts. Then, in \autoref{sec:transform}, we present our proposed \emph{forward} transformation and the accompanying verification workflow, as well as the theory behind proof- and counterexample-generation. Finally, in \autoref{sec:evaluation}, we present our experimental results comparing the effect of using the existing \emph{backward} transformation versus the novel \emph{forward} transformation on the performance of the verification workflow.
\section{Background}
\label{sec:background}

In this section, we introduce the theoretical background for the paper, including \emph{software verification}, \emph{control flow automata} (CFAs), and \emph{Counterexample-Guided Abstraction Refinement} (CEGAR).

\subsection{Formal Software Verification}

The goal of software verification is to mathematically prove certain properties of a program. One such property is the reachability of labelled control locations. A program is \emph{unsafe} if such a location can be reached from the initial location of the program using a finite number of transitions; otherwise, it is \emph{safe}. Due to the uncertainties and complexity of dealing with high-level programming languages, the input is first transformed into a formal representation \cite{formalise}. \emph{Model checking} is then often employed \cite{clarke_grumberg_peled_1999}, which explores the state space of the program, thus verifying the reachability of the error states. While generally this problem is undecidable \cite{turing1936computable}, and enumerating the state space naively is infeasible in practice \cite{clarke_klieber_novacek_zuliani_2012}, there exist efficient algorithms for solving a subset of the input tasks, such as the Counterexample-Guided Abstraction Refinement (CEGAR) technique \cite{cegar}.  

\subsubsection{Control Flow Automata}

A \emph{Control Flow Automaton} represents a program as a directed graph. Formally, a control flow automaton is a tuple ${CFA} = (V, L, l_0, E)$, where:
\begin{itemize}\setlength\itemsep{0.1em}
    \item $V$: A set of \emph{variables}, where each $v \in V$ can have values from its domain $D_v$.
    \item $L$: A set of \emph{locations}, where each \emph{location} can be interpreted as a possible value of the program counter.
    \item $l_0 \in L$: The \emph{initial location}, that is active at the start of the program.
    \item $E \subseteq L \times Ops \times L $: A set of transitions, where a transition is a directed edge going from one location in $L$ to another, with a label $op \in Ops$, where $Ops$ is a set of operations that can be executed as the program advances from one location to another. An $op \in Ops$ can be one of the following:
\begin{itemize}\setlength\itemsep{0.1em}
    \item $v = expr$: An assignment of a variable, where the value of $v \in V$ becomes the evaluation of the right-hand side $expr$. 
    \item $havoc \; v$: A non-deterministic assignment of a variable, after which the value of $v \in V$ can be anything from its domain $D_v$.
    \item $[cond]$: A \emph{guard} operation, where \emph{cond} is an expression that evaluates to a boolean value. The transition can only be executed if the \emph{cond} in the \emph{guard} evaluates to $true$.
\end{itemize}
\end{itemize} 

In formal software verification, it is also useful to distinguish \emph{error locations}, which are locations where the program would behave in an undesirable way, as well as \emph{final locations}, which have no \emph{outgoing transitions}.

The representation of program execution on the CFA consists of an alternating sequence of locations and operations, where at each location, the \emph{state} of the CFA can be described as $S = (l_S, d_1, d_2, ..., d_n)$, where:
\begin{itemize}\setlength\itemsep{0.1em}
    \item $l \in L$ is the current location of the program,
    \item $d_1, d_2, ..., d_n$ are the values of all variables, that is $v_i = d_i, v_i \in V, d_i \in D_{v_i}$, for every $1 \leq i \leq |V|$.
\end{itemize}

All possible states of the CFA make up the \emph{state space} of the program. The operations in an alternating sequence (representing an execution of the program) can then be interpreted as \emph{transitions} in the state-space of the program.

\subsection{Counterexample-Guided Abstraction Refinement}

\begin{figure}
\begin{minipage}[b]{0.66\textwidth}
\begin{figure}[H]
    \centering
    \begin{tikzpicture}[scale=1, every node/.style={transform shape}]
    \tikzset{vertex/.style = {shape=ellipse, draw, minimum size=1.5em}}
    \node[vertex,draw=none] (Initprec) at  (0,0) {Initial precision};
    \node[vertex,shape=rectangle] (Abs) at  (0,-1.5) {Abstractor};
    \node[vertex,shape=rectangle] (Ref) at  (7,-1.5) {Refiner};
    \node[vertex] (ARG) at  (3.5,-1.5) {ARG};
    \node[vertex] (Safe) at  (0,-3) {Safe};
    \node[vertex] (Unsafe) at  (7,-3) {Unsafe};
    
    \tikzset{edge/.style = {->,-{Latex[scale=1.5]},color=black}}
    \draw[edge] (Initprec) to (Abs);
    \draw[edge] (Abs) to (Safe);
    \draw[edge] (Ref) to (Unsafe);
    \draw[edge] (Abs) [bend left=31] to node [midway, above] {Abstract counterexample} (Ref);
    \draw[edge] (Ref) [bend left=31] to node [midway, below] {Refined precision} (Abs);
    \draw[edge,dashed] (Abs) to node [midway, above] {Expand} (ARG);
    \draw[edge,dashed] (Ref) to node [midway, above] {Prune} (ARG);
\end{tikzpicture}
    \caption{The CEGAR loop}
    \label{fig:cegarloop}
\end{figure}
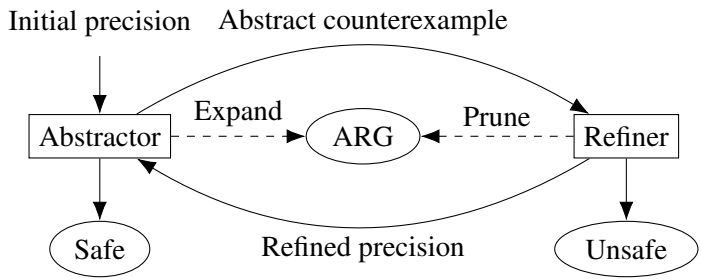
\end{minipage}%
\begin{minipage}[b]{0.33\textwidth}
\begin{figure}[H]
    \centering
    \includegraphics[width=0.7\textwidth]{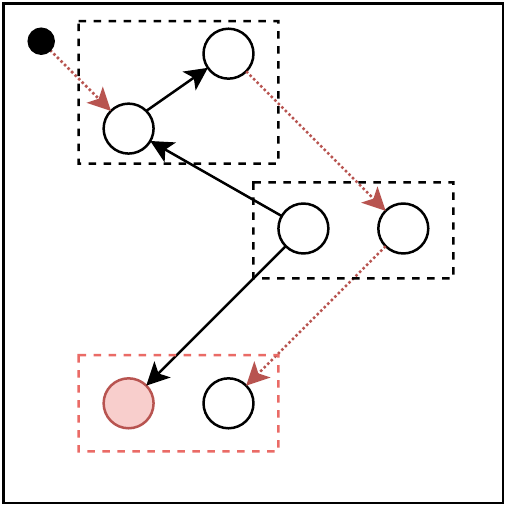}
    \caption{Abstract state space}
    \label{fig:arg}
\end{figure}
\end{minipage}
\end{figure}

\emph{Counterexample-Guided Abstraction Refinement (CEGAR)} \cite{cegar} is an abstraction-based model checking algorithm.%

The core of the algorithm is the CEGAR-loop (\autoref{fig:cegarloop}), made up of two main parts: the \emph{abstractor} and the \emph{refiner}. The abstractor builds the \emph{Abstract Reachability Graph} (ARG, a directed and acyclic graph containing abstract states and interconnecting transitions) using the \emph{expand} operation and \emph{covering} relation on abstract states. A parameter of abstraction is precision, which describes how much information about a concrete state is abstracted in the abstract state. An abstract state is an overapproximation of the possible concrete states (as seen in \autoref{fig:arg}), consequently, if no abstract error-state is reachable, then no concrete error-state is reachable, meaning the program is \emph{safe}.

On the other hand, if an abstract error-state is reachable, the abstractor produces an \emph{abstract counterexample}, starting at the initial abstract state and ending in an abstract error state. The refiner then decides whether a concrete error state is reachable in the abstract error state. If it can be reached, then the program is \emph{unsafe}, and the path from the initial location of the CFA to a concrete error state is presented as a counterexample.

However, if a concrete error-state is not reachable, then the reachability of the abstract error state is a result of the overapproximation of abstraction, as demonstrated in \autoref{fig:arg}. Thus, the abstraction needs to be \emph{refined} so that the abstract error state does not contain the unreachable concrete error state. This results in a refined precision, which is passed back to the abstractor after all unreachable abstract states are removed (\emph{pruned}) from the abstract state-space.

The CEGAR loop is repeated until it either finds a concrete counterexample to the safety of the program or proves that no abstract error-state is reachable, that is, all nodes in the ARG are either expanded or covered. In the first case, the program is \emph{unsafe}, while in the latter, it is \emph{safe}.

\section{Transforming Constrained Horn Clauses to Control Flow Automata}
\label{sec:transform}

\newcounter{step}[subsection]
\newcommand{\stepautorefname}{Step}
\newenvironment{step}[1][]{\refstepcounter{step}\par\medskip
\noindent\textbf{Step~\thestep. #1} \rmfamily}{\medskip}
\newcounter{altstep}[step]
\def\thealtstep{\thestep/\alph{altstep}}
\newcommand{\altstepautorefname}{Step}
\newenvironment{altstep}[1][]{\refstepcounter{altstep}\par\medskip
\noindent\textbf{Step~\thealtstep. #1} \rmfamily}{\medskip}

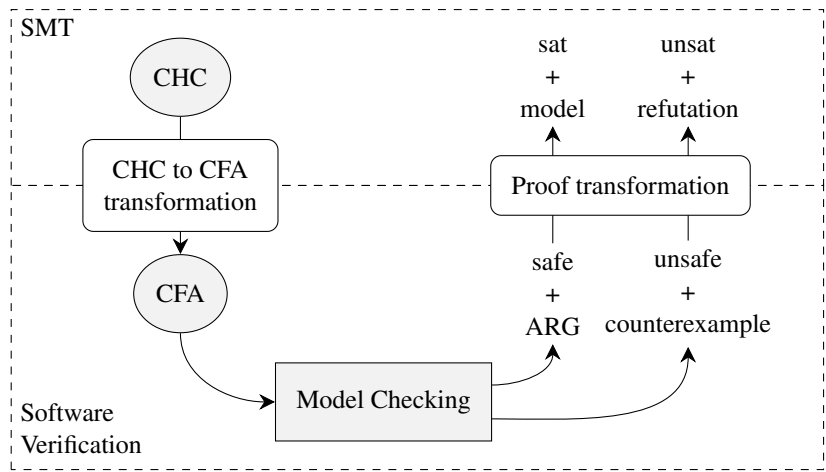
\begin{figure}[!ht]
    \centering
    \begin{tikzpicture}[scale=0.9, every node/.style={transform shape}]
    \draw[dashed] (0,0.8) rectangle (12,5);
    \node[above right,align=left] at (0,0.9) {Software \\ Verification};
    \draw[dashed] (0,5) rectangle (12,7.6);
    \node[below right] at (0,7.6) {SMT};

    \tikzset{vertex/.style = {draw, fill=gray!10, minimum height=3em}}
    \tikzset{t/.style = {align=center}}
    \node[vertex, ellipse] (chc) at (2.5, 6.6) {CHC};
    \node[vertex, ellipse] (cfa) at (2.5, 3.4) {CFA};
    \node[vertex, rectangle, inner sep=0.8em] (mc) at (5.5, 1.8) {Model Checking};
    \node[t] (safe) at (8, 3.4) {safe \\ + \\ ARG};
    \node[t] (unsafe) at (10, 3.4) {unsafe \\ + \\ counterexample};
    \node[t] (sat) at (8, 6.6) {sat \\ + \\ model};
    \node[t] (unsat) at (10, 6.6) {unsat \\ + \\ refutation};

    \tikzset{edge/.style = {->,-{Stealth[width=6pt, length=6pt]},color=black}}
    \draw[edge] (chc) to (cfa);
    \draw[edge] (cfa.south) [out=270, in=180] to (mc.west);
    \draw[edge] (mc.east) +(0,0.25) [out=0, in=270] to (safe.south);
    \draw[edge] (mc.east) +(0,-0.25) [out=0, in=270] to (unsafe.south);
    \draw[edge] (safe) to (sat);
    \draw[edge] (unsafe) to (unsat);
    
    \tikzset{board/.style = {draw, rectangle, rounded corners, fill=white, align=center, inner sep=0.8em}}
    \node[board] at (2.5,5) {CHC to CFA\\transformation};
    \node[board] at (9,5) {Proof transformation};
\end{tikzpicture}
	\caption{Overview of the presented work.}
	\label{fig:overview}
\end{figure}

In this section, we present a novel approach of CHC to CFA transformation. The goal of this transformation is to create a CFA from a linear CHC in a way that turns the SMT problem of satisfiability in a CHC into a software verification question of erroneous state reachability in the CFA, so that model checking techniques can be used to decide both. More specifically, an erroneous state in a CFA should be reachable if, and only if the CHC is unsatisfiable. In this case, a refutation of the satisfiability should be given; otherwise a satisfying model ought to be generated. The approach is summarized in \autoref{fig:overview}.

The transformation consists of two parts: the mapping of CHCs to CFAs, and the generation of a model/refutation from the output of model checking. These are represented in \autoref{fig:overview} by the boxes \emph{CHC to CFA transformation} and \emph{Proof transformation}, respectively, and are not to be confused with \emph{forward} and \emph{backward} transformations described later on. As seen in the figure, proof transformation requires the utilized model checking algorithm to provide a counterexample when the CFA is deemed unsafe, and to produce an ARG when the CFA is safe.

The main idea behind the CHC to CFA transformation is to represent the uninterpreted functions as locations in the CFA, map CHCs to edges guarded by the conditions in the CHC, and use local variables to model the implications of deductions. The deducibility of a predicate with certain parameters can then be represented by the corresponding location's reachability during verification, with the given parameters as the variables' values. The source of the edges of fact CHCs can be the initial location of a CFA, since these do not have any preconditional predicates in their bodies. The target of the edge of a query CHC can then be an error location, which can only be reached if the conditions on an incoming edge are satisfied, similarly to how $\bot$ is deduced. If the error location can be reached from the initial location, then the counterexample contains the path of edges to it, which can then be mapped to their CHCs to show a sequence of CHCs that deduce $\bot$ from facts. On the other hand, if the error location is unreachable, then the explored abstract states can be used to define the uninterpreted functions to provide a satisfying model.

One way of approaching the problem of CHC satisfiability is to start with the facts, and try to apply the induction and query CHCs to deduce $\bot$. This is called the \emph{forward} or \emph{bottom-up} approach, which is what our main contribution, the \emph{forward transformation} employs. Another approach is to recursively check what would be required to satisfy the body of the query CHC, stopping only when all requirements are satisfied by facts. We refer to this as a \emph{backward} or \emph{top-down} approach, which is used by \textsc{Ultimate Unihorn} \cite{chc-comp_report22} to transform CHCs into program code.

An example CHC problem will be used throughout the chapter to demonstrate the transformations.

\begin{example}
\label{ex:mot}
Consider the following CHC problem within integer arithmetic:

\begin{align}
    A(n) &\leftarrow n > 0 \land n < 100 \label{eq:me_f1} \\ 
    B(n, x) &\leftarrow A(n) \land x > 0 \label{eq:me_i1} \\
    C(y, x) &\leftarrow B(n, x) \land y = n - x \land y > 0 \label{eq:me_i2} \\
    A(n) &\leftarrow C(y, x) \land n = y + (y \bmod x) \label{eq:me_i3} \\
    \bot &\leftarrow A(n) \land n \geq 100 \label{eq:me_q1}
\end{align}

The fact states that $A(n)$ needs to evaluate to true for $0 < n < 100$, while the satisfiability of the query depends on $A(n)$ being false for $n \geq 100$ and $n \leq 0$. What makes this problem non-trivial is the cyclic deductions between the predicates $A, B$ and $C$: $B$ can be deduced from $A$, $C$ can be deduced from $B$, and $A$ can be deduced from $C$ under certain conditions. Trying a naive, manual deduction approach becomes a bit cumbersome here, due to the possibility of an infinite deduction cycle and the high number of combinations possible between the variables' values. %

One may notice that $n$ can not increase in the cycle since no matter what the subtracted $x$ is, it will always be larger than the $y \bmod x$ that is added to $n$ in a cycle. In the following, it will be shown that the problem is indeed satisfiable, by transforming it into a software verification problem and synthesizing a satisfying model from its proof. The CFA resulting from the transformation can be seen on \autoref{fig:forward_ex}. The effect of each step on the CFA is explained as the steps are introduced.

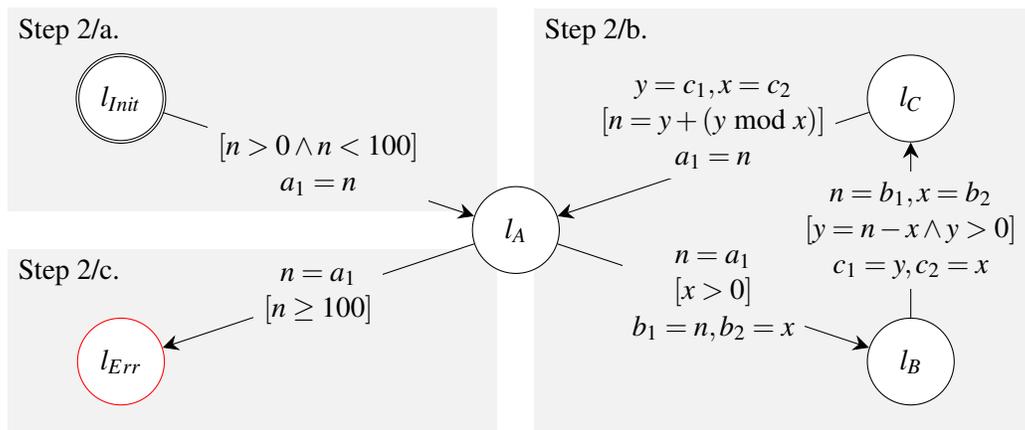
\begin{figure}[!ht]
	\centering
	\begin{tikzpicture}[every node/.style={transform shape}]
    \tikzset{area/.style = {fill=gray!10}}
    \fill[area] (-1.5,5.5) rectangle (5,8.2);
    \node[below right] at (-1.5,8.2) {\autoref{st:f2a}.};
    \fill[area] (5.5,2.5) rectangle (12.2,8.2);
    \node[below right] at (5.5,8.2) {\autoref{st:f2b}.};
    \fill[area] (-1.5,2.5) rectangle (5,5);
    \node[below right] at (-1.5,5) {\autoref{st:f2c}.};
    
    \tikzset{vertex/.style = {draw, circle, fill=white, minimum size=3em}}
    \node[vertex, double] (init) at (0, 7) {$l_{Init}$};
    \node[vertex, draw=red] (err) at (0, 3.5) {$l_{Err}$};
    \node[vertex] (A) at (5.25, 5.25) {$l_A$};
    \node[vertex] (B) at (10.5, 3.5) {$l_B$};
    \node[vertex] (C) at (10.5, 7) {$l_C$};

    \tikzset{edge/.style = {->,-{Stealth[width=6pt, length=6pt]},color=black}}
    \tikzset{label/.style = {align=center, area, midway}}
    \draw[edge] (init) to node[label] {$[n > 0 \land n < 100]$ \\ $a_1 = n$} (A);
    \draw[edge] (A) to node[label] {$n = a_1$ \\ $[n \geq 100]$} (err);
    \draw[edge] (A) to node[label] {$n = a_1$ \\ $[x > 0]$ \\ $b_1 = n, b_2 = x$} (B);
    \draw[edge] (B) to node[label] {$n = b_1, x = b_2$ \\ $[y = n - x \land y > 0]$ \\ $c_1 = y, c_2 = x$} (C);
    \draw[edge] (C) to node[label,yshift=15pt] {$y = c_1, x = c_2$ \\ $[n = y + (y \bmod x)]$ \\ $a_1 = n$} (A);
\end{tikzpicture}
	\caption{CFA of \autoref{ex:mot} after forward transformation.}
	\label{fig:forward_ex}
\end{figure}

\end{example}

\subsection{Constrained Horn Clause Transformation}
\label{sec:forward_t}

The transformation first creates the locations and variables of the CFA, then maps the CHCs to edges in different ways for fact, induction and query CHCs.

Consider the linear CHC problem with CHC set $\{C_1, C_2, \dots, C_k\}$ over the following uninterpreted functions:
\begin{align*}
    B_1(b_1^1, b_2^1, \dots, b_{m_1}^1), B_2(b_1^2, b_2^2, \dots, b_{m_2}^2), \dots,  B_n(b_1^n, b_2^n, \dots, b_{m_n}^n)
\end{align*}
That is, each CHC $C_l, \forall l \in \{1, 2, \dots, k\}$ takes one of the following three forms for some $i, j \in \{1, 2, \dots, k\}$:
\begin{align*}
    B_i(x_1, x_2, \dots, x_{m_i}) &\leftarrow \varphi_l, \\
    B_i(x_1, x_2, \dots, x_{m_i}) &\leftarrow B_j(y_1, y_2, \dots, y_{m_j}) \land \varphi_l, \\
    \bot &\leftarrow B_j(y_1, y_2, \dots, y_{m_j}) \land \varphi_l,
\end{align*}
where $\varphi_l$ is the interpreted formula in the body of $C_l$. As before, CHCs in these forms are referred to as facts, inductions and queries, respectively.

\begin{step}[Create CFA locations and variables]
\label{st:f1}

The uninterpreted functions $B_1(b_1^1, b_2^1, \dots, b_{m_1}^1)$, $\dots$,  $B_n(b_1^n, b_2^n, \dots, b_{m_n}^n)$ are mapped to the $CFA = (V, L, l_{Init}, E)$, where:

\begin{itemize}\setlength\itemsep{0.1em}
    \item $V = \{b_j^i \, | \, \forall i \in \{1, 2, \dots, n\} : \forall j \in \{1, 2, \dots, m_{i}\} \}$,
    \item $L = \{l_{Init}, l_{Err}, l_1, l_2, \dots, l_n\}$,
    \item $l_{Init}$,
    \item $E = \varnothing$.
\end{itemize}

Semantically, a new location is created for each uninterpreted function, along with an initial location $l_{Init}$ and a distinguished error location $l_{Err}$. In addition, a unique variable is created for each parameter in every predicate. It is worth noting that the edge set is empty at this point, because edges are added in the next step of the transformation.

\end{step}

The motivation behind creating a location and variables for every uninterpreted function is that this way, a location's reachability with certain variable values can be directly mapped to the predicate's evaluation with said variable values as parameters: if a location $l_i$ representing $C_i$ is reachable with some values for variables $b_1^i, b_2^i, \dots, b_{m_i}^i$, then $C_i(b_1^i, b_2^i, \dots, b_{m_i}^i)$ should evaluate to true. On the other hand, if $l_i$ can not be reached with variables $b_1^i, b_2^i, \dots, b_{m_i}^i$, then $C_i(b_1^i, b_2^i, \dots, b_{m_i}^i)$ ought to evaluate to false.

\begin{example}
\label{ex:fw_cfa_init}
From \autoref{ex:mot}, the first step of the forward transformation would create the $CFA = (V, L, l_{Init}, \varnothing)$, with locations $L = \{l_{Init}, l_{Err}, l_A, l_B, l_C\}$ and variables $V = \{a_1, b_1, b_2, c_1, c_2\}$. The created locations can be seen in white on the CFA in \autoref{fig:forward_ex}.
\end{example}

\begin{step}[Create CFA edges]
\label{st:f2}

In this step, each CHC is transformed into an edge in the CFA created in \autoref{st:f1}. Each kind of CHC (fact, induction, query) is treated differently, as described in the following subsections. The goal of this mapping is for the transition on the edge to only be possible, when the head of the CHC is deducible from the body of it.

\begin{altstep}[Create fact edges]
\label{st:f2a}

For each fact CHC $C_l: B_i(x_1, x_2, \dots, x_{m_i}) \leftarrow \varphi_l$ where $i \in \{1, 2, \dots, n\}$, an edge is created from the initial location $l_{Init}$ to $l_i$, the location representing $B_i$. The labels on the created edge consist of the following, in the specified order:

\begin{itemize}\setlength\itemsep{0.1em}
    \item $\varphi_l$, the interpreted formula in the CHC's body as a guard,
    \item $b_1^i = x_1, b_2^i = x_2, \dots, b_{m_i}^i = x_{m_i}$, assignment of the passed values to the variables corresponding to the input parameters.
\end{itemize}

Fact CHCs are named facts because they can be deduced just from the background theory $\top$, when the interpreted formula $\varphi_l$ is true. The created edge from the initial location mimics this, since the target of an edge will be reachable from the initial location when the guard $\varphi$ is true.

\end{altstep}

To put it more formally, the head of a fact CHC $B_i(x_1, x_2, \dots, x_{m_i})$ is only deducible when its body, the interpreted formula $\varphi_l$ is true. Similarly, the location $l_i$ is only reachable from the initial location $l_{Init}$ of the CFA using the created edge, when its guard $\varphi_l$ evaluates to true. Furthermore, the parameters $x_1, x_2, \dots, x_{m_i}$ are assigned to $b_1^i, b_2^i, \dots, b_{m_i}^i$, meaning that the constraints of $\varphi_l$ on the parameters are applied to the variables related to the location, just as they are applied when deducing $B_i(x_1, x_2, \dots, x_{m_i})$. Thus, we can conclude that $l_i$ is only reachable using the created edge with variables $b_1^i, b_2^i, \dots, b_{m_i}^i$ valued $x_1, x_2, \dots, x_{m_i}$, when $B_i(x_1, x_2, \dots, x_{m_i})$ is deducible using $C_l$.

\begin{example}
In \autoref{ex:mot}, the second step of the forward transformation for fact CHCs would create the edge $e = (l_{Init}, op, l_A)$ from \autoref{eq:me_f1}, where the guard of $op$ would be $n > 0 \land n < 100$, and the assignments would consist of $a_1 = n$, since $a_1$ is the variable corresponding to the first (and only) parameter of the predicate $A$. The created edge can be seen in the top-left gray rectangle on the CFA in \autoref{fig:forward_ex}.
\end{example}

\begin{altstep}[Create induction edges]
\label{st:f2b}

For each induction CHC $C_l: B_i(x_1, x_2, \dots, x_{m_i}) \leftarrow B_j(y_1, y_2, \dots, y_{m_j}) \land \varphi_l$ where $i, j \in \{1, 2, \dots, n\}$, an edge is created from $l_j$ (the location representing $B_j$) to $l_i$ (the location representing $B_i$). The labels on the created edge consist of the following, in the specified order:

\begin{itemize}\setlength\itemsep{0.1em}
    \item $y_1 = b_1^j, y_2 = b_2^j, \dots, y_{m_j} = b_{m_j}^j$, assignment of the variables corresponding to the input parameters of $B_j$ to the passed values,
    \item $\varphi_l$, the interpreted formula in the CHC's body as a guard,
    \item $b_1^i = x_1, b_2^i = x_2, \dots, b_{m_i}^i = x_{m_i}$, assignment of the passed values to the variables corresponding to the input parameters of $B_i$.
\end{itemize}

In addition to the first assignments, $x_1, x_2, \dots, x_{m_i}$ and all variables in $\varphi_l$ need to be uninitialized with a $havoc$ statement to ensure that the semantics of $\forall$ in the CHCs are kept. However, the $havoc$ statements are omitted from the examples for ease of readability. The order of instructions is also important: the assignments from the source location's variables need to happen before $\varphi_l$ is evaluated.

Induction CHCs embody deductions from their bodies to their heads with some conditions $\varphi_l$. Assuming that $l_j$ could have only been reached if it is deducible with some parameters, then this edge resembles the same: one can only go to $l_i$ from $l_j$ when $\varphi_l$ is true.

\end{altstep}

More formally, the head of an induction CHC $B_i(x_1, x_2, \dots, x_{m_i})$ is only deducible, when $\varphi_l$ is true and  $B_j(y_1, y_2, \dots, y_{m_j})$ is deducible. Similarly, the location $l_i$ can only be reached from $l_j$ once $l_j$ has been reached and the guard $\varphi_l$ evaluates to true. Furthermore, the variables $b_1^j, b_2^j, \dots, b_{m_j}^j$ are assigned to $y_1, y_2, \dots, y_{m_j}$ and the parameters $x_1, x_2, \dots, x_{m_i}$ are assigned to $b_1^i, b_2^i, \dots, b_{m_i}^i$, meaning that the constraints of $\varphi_l$ are applied to the $y$ parameters and the $b^i$ variables related to the location $l_i$, just as they are applied when deducing $B_i(x_1, x_2, \dots, x_{m_i})$ from $B_j(y_1, y_2, \dots, y_{m_j})$. Thus, we can conclude that $l_i$ is only reachable using the created edge with variables $b_1^i, b_2^i, \dots, b_{m_i}^i$ valued $x_1, x_2, \dots, x_{m_i}$ from $l_j$ with variables $b_1^j, b_2^j, \dots, b_{m_j}^j$ valued $y_1, y_2, \dots, y_{m_j}$, when $B_i(x_1, x_2, \dots, x_{m_i})$ is deducible from $B_j(y_1, y_2, \dots, y_{m_j})$ using $C_l$.

\begin{example}
From \autoref{ex:mot}, the second step of the forward transformation for induction CHCs would create three edges from \autoref{eq:me_i1}, \ref{eq:me_i2} and \ref{eq:me_i3}:

\begin{itemize}\setlength\itemsep{0.1em}
    \item $e_1 = (l_A, op_1, l_B)$ for $B(n, x) \leftarrow A(n) \land x > 0$, where $op_1$ consists of the assignment $n = a_1$, then the guard $x > 0$, and the assignments $b_1 = n, b_2 = x$ at last,
    \item $e_2 = (l_B, op_2, l_C)$ for $C(y, x) \leftarrow B(n, x) \land y = n - x \land y > 0$, where $op_2$ consists of the assignments $n = b_1, x = b_2$, then the guard $y = n - x \land y > 0$, and the assignments $c_1 = y, c_2 = x$ at last,
    \item $e_3 = (l_C, op_3, l_A)$ for $A(n) \leftarrow C(y, x) \land n = y + (y \bmod x)$, where $op_3$ consists of the assignments $y = c_1, x = c_2$, then the guard $n = y + (y \bmod x)$, and the assignment $a_1 = n$ at last.
\end{itemize}

The created edges can be seen in the right-hand side gray rectangle on the CFA in \autoref{fig:forward_ex}.
\end{example}

\begin{altstep}[Create query edges]
\label{st:f2c}

For each query CHC $C_l: \bot \leftarrow B_j(y_1, y_2, \dots, y_{m_j}) \land \varphi_l$ where $j \in \{1, 2, \dots, n\}$ an edge is created to the error location $l_{Err}$ from $l_j$, the location representing $B_j$. The labels on the created edge consist of the following, in the specified order:
\setlength\itemsep{0.1em}
\begin{itemize}
    \item $y_1 = b_1^j, y_2 = b_2^j, \dots, y_{m_j} = b_{m_j}^j$, assignment of the variables corresponding to the input parameters to the passed values,
    \item $\varphi_l$, the interpreted formula in the CHC's body as a guard.
\end{itemize}

The bodies of CHC queries should not be deducible, otherwise $\bot$ can be deduced and the problem is unsatisfiable. This behaviour is captured by the created edge: if the edge's source is reachable with values that make the guard of the edge true, then the error location is reachable, making the program unsafe.

\end{altstep}

In a formal way, the head of the query CHC $\bot$ is only deducible when both $B_j(y_1, y_2, \dots, y_{m_j})$ is deducible, and $\varphi_l$ is true. Similarly, the error location $l_{Err}$ can only be reached from $l_j$ once $l_j$ has been reached and the guard $\varphi_l$ evaluates to true. Furthermore, the variables $b_1^j, b_2^j, \dots, b_{m_j}^j$ are assigned to $y_1, y_2, \dots, y_{m_j}$, meaning that the constraints of $\varphi_l$ are applied to the $y$ parameters, just as they are applied when deducing $\bot$ from $B_j(y_1, y_2, \dots, y_{m_j})$. Thus, we can conclude that $l_{Err}$ is only reachable using the created edge from $l_j$ with variables $b_1^j, b_2^j, \dots, b_{m_j}^j$ valued $y_1, y_2, \dots, y_{m_j}$, when $\bot$ is deducible from $B_j(y_1, y_2, \dots, y_{m_j})$ using $C_l$.

\begin{example}
In \autoref{ex:mot}, the second step of the forward transformation for query CHCs would create the edge $e = (l_A, op, l_{Err})$ from \autoref{eq:me_q1}, where op would consist of the assignment $n = a_1$ and the guard $n \geq 100$. The created edge can be seen in the bottom-left gray rectangle on the CFA in \autoref{fig:forward_ex}.
\end{example}

\end{step}

To summarize, first a location $l_i$ and variables $b_1^i, b_2^i, \dots, b_{m_i}^i$ are created for each uninterpreted function $B_i(b_1^i, b_2^i, \dots, b_{m_i}^i)$, then all CHCs are transformed into edges. Since the edges are created in a way that $l_i$ can only be reached with the corresponding variables $b_1^i, b_2^i, \dots, b_{m_i}^i$ valued $x_1, x_2, \dots, x_{m_i}$ if, and only if $B_i(x_1, x_2, \dots, x_{m_i})$ can be deduced, we can conclude that the described transformation successfully converts the problem of satisfiability into a question of error location reachability. Thus, using a model checker to decide the latter will yield a result for the former as well: if the CFA is \emph{unsafe}, the CHC problem is \emph{unsatisfiable}; if the CFA is \emph{safe}, the CHC problem is \emph{satisfiable}.

It is worth to consider what the transformation results in, when there is no fact or query CHC in the set of CHCs. In the former case, there will not be any outgoing edges from the initial location of the CFA. As a result, none of the locations will be reachable, meaning the predicates need not be true for any input, which can be expressed as $B_i \equiv false, \forall i \in \{1, 2, \dots, n\}$.

In the latter case, there will not be any edges going to the error location of the CFA. As a result, all locations are reachable in the abstract state $\top$, meaning the predicates can be true for any input, which can be expressed as $B_i \equiv true, \forall i \in \{1, 2, \dots, n\}$.

\subsection{Proof Transformation}

Proof transformation is the step of converting the result of the model checking algorithm to an answer to the CHC problem. This consists of two parts, depending on the result: the generation of a satisfying model from the ARG built during verification, or the creation of a refutation from the counterexample provided by the model checking algorithm.

\subsubsection{Satisfying Model Generation}
\label{sec:forward_model}

An SMT problem is called \emph{satisfiable}, when a \emph{model} (i.e., an assignment to constants) fulfilling all constraints exists. In the case of a CHC problem this means the definition of all uninterpreted functions $B_1(b_1^1, b_2^1, \dots, b_{m_1}^1)$, $B_2(b_1^2, b_2^2, \dots, b_{m_2}^2)$, $\dots$,  $B_n(b_1^n, b_2^n, \dots, b_{m_n}^n)$ present in the set of CHCs, that satisfy all of the CHCs.

The transformation in \autoref{sec:forward_t} ensures that a location $l_i$ in the CFA can only be reached with the corresponding variables $b_1^i, b_2^i, \dots, b_{m_i}^i$ valued $x_1, x_2, \dots, x_{m_i}$ if, and only if $B_i(x_1, x_2, \dots, x_{m_i})$ can be deduced. If a node $S_j = (l_i, L_1^j, \dots, L_{k_j}^j)$ is present in the ARG, it means $l_i$ has been reached under the condition $L_1^j \land \dots \land L_{k_j}^j$. Consequently, it is guaranteed that $B_i$ can be deducted under the condition $L_1^j \land \dots \land L_{k_j}^j$. This is true for all $S^i = \{S_j \, | \, S_j = (l_i, L_1^j, \dots, L_{k_j}^j)\}$ nodes in the ARG, therefore $B_i$ needs to evaluate to true under either of their conditions, which can be represented by concatenating them with $\lor$. This gives the following the definition for $B_i, \forall i \in \{1, 2, \dots, n\}$:

\begin{align}
\label{eq:forward_def}
    B_i(b_1^i, b_2^i, \dots, b_{m_i}^i) =
    \bigvee_{S_j = (l_i, L_1^j, \dots, L_{k_j}^j)}^{S^i} \left( L_1^j \land \dots \land L_{k_j}^j \right)
\end{align}

At the end of verification of a safe CFA, the ARG is fully expanded, i.e., all reachable abstract states have been visited and none are in an erroneous location. Furthermore, no erroneous state can be reached from any of the nodes in the ARG. Therefore the definitions provided by \autoref{eq:forward_def} guarantee that there can not be a deduction to $\bot$, meaning they satisfy the CHC problem.

The type of information present in any $L^j$ needs to be taken into consideration when defining the function. If $L^j$ contains information about any other variable $x$ then the variables $b_1^i, b_2^i, \dots, b_{m_i}^i$ representing the input parameters of $B_i$, then unless some information about a $b^i$ is dependent on $x$ (e.g. $b_1^i > x$), $L^j$ can be left out. If there is a dependent $b^i$, then $x$ needs to be defined with a universal quantifier inside the function ($\forall x$).

\begin{example}
Applying model checking with predicate abstraction to the CFA in \autoref{fig:forward_ex} may result in the Abstract Reachability Graph (ARG) seen in \autoref{fig:forward_ARG}. The regular arrows represent transitions between abstract states, while the dotted arrow denotes that the source abstract state is covered by the target abstract state.

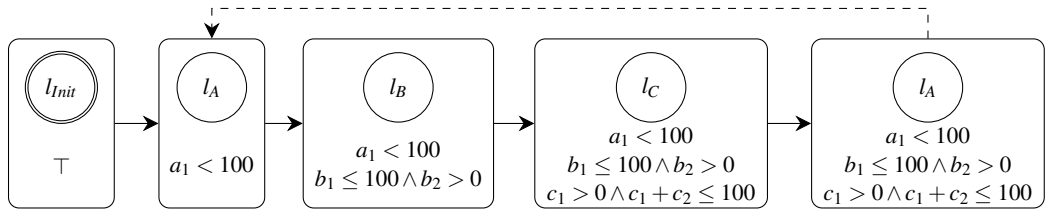
\begin{figure}[!ht]
	\centering
	\begin{tikzpicture}[scale=0.8, every node/.style={transform shape}]
    \tikzset{vertex/.style = {draw, circle, fill=white, minimum size=3em}}
    \tikzset{state/.style = {draw, rectangle, rounded corners}}
    \tikzset{label/.style = {align=center}}

    \node[state, minimum width=50, minimum height=80] (0) at (0, 0) {};
    \node[vertex, double] (init) at (0, 0.6) {$l_{Init}$};
    \node[label] at (0, -0.7) {$\top$};

    \node[state, minimum width=50, minimum height=80] (1) at (2.5, 0) {};
    \node[vertex] (A1) at (2.5, 0.6) {$l_A$};
    \node[label] at (2.5, -0.7) {$a_1 < 100$};
    
    \node[state, minimum width=90, minimum height=80] (2) at (5.6, 0) {};
    \node[vertex] (B) at (5.6, 0.6) {$l_B$};
    \node[label] at (5.6, -0.7) {$a_1 < 100$\\$b_1 \leq 100 \land b_2 > 0$};

    \node[state, minimum width=110, minimum height=80] (3) at (9.8, 0) {};
    \node[vertex] (C) at (9.8, 0.6) {$l_C$};
    \node[label] at (9.8, -0.7) {$a_1 < 100$\\$b_1 \leq 100 \land b_2 > 0$\\$c_1 > 0 \land c_1 + c_2 \leq 100$};

    \node[state, minimum width=110, minimum height=80] (4) at (14.4, 0) {};
    \node[vertex] (A2) at (14.4, 0.6) {$l_A$};
    \node[label] at (14.4, -0.7) {$a_1 < 100$\\$b_1 \leq 100 \land b_2 > 0$\\$c_1 > 0 \land c_1 + c_2 \leq 100$};
    
    \tikzset{edge/.style = {->,-{Stealth[width=6pt, length=6pt]},color=black}}
    \draw[edge] (0) to node[label] {} (1);
    \draw[edge] (1) to node[label] {} (2);
    \draw[edge] (2) to node[label] {} (3);
    \draw[edge] (3) to node[label] {} (4);
    \draw[edge, dashed] (4.north) -- ($(4.north)+(0,0.5)$) -- ($(1.north)+(0,0.5)$) -- (1.north);
\end{tikzpicture}
	\caption{ARG resulting from the model checking of the CFA in \autoref{fig:forward_ex}.}
	\label{fig:forward_ARG}
\end{figure}

As described in \autoref{ex:fw_cfa_init}, the uninterpreted function $A(n)$ corresponds to the location $l_A$ and the variable $a_1$. Therefore its definition depends on the predicates of the abstract states that are in $l_A$, more specifically $(l_A, a_1 < 100)$ and $(l_A, a_1 < 100 \land b_1 \leq 100 \land b_2 > 0 \land c_1 > 0 \land c_1 + c_2 \leq 100)$. Using these states, we can define $A(n)$ as the disjunction of the predicates by converting $a_i$ to $n$: $A(n) = n < 100 \lor (n < 100 \land b_1 \leq 100 \land b_2 > 0 \land c_1 > 0 \land c_1 + c_2 \leq 100), \forall b_1, b_2, c_1, c_2$. Since predicates of $n$ do not depend on other variables, they can be left out, leading to $A(n) = n < 100 \lor n < 100 = n < 100$.

Similarly, $B(n, x)$ can be defined using abstract states that are in $l_B$, namely the single abstract state $(l_B, a_1 < 100 \land b_1 \leq 100 \land b_2 > 0)$. Converting $b_1$ and $b_2$ back to $n$ and $x$ gives $B(n, x) = a_1 < 100 \land n \leq 100 \land x > 0, \forall a_1$, which can also be simplified to $B(n, x) = n \leq 100 \land x > 0$ by omitting unused variables.

Lastly, $C(y, x)$ is defined using the abstract state $(l_C, a_1 < 100 \land b_1 \leq 100 \land b_2 > 0 \land c_1 > 0 \land c_1 + c_2 \leq 100)$. Converting $c_1$ and $c_2$ back to $y$ and $x$ results in $C(y, x) = a_1 < 100 \land b_1 \leq 100 \land b_2 > 0 \land y > 0 \land y + x \leq 100, \forall a_1, b_1, b_2$, which leads to the definition of $C(y, x) = y > 0 \land y + x \leq 100$ after getting rid of unused variables.

While it may not be trivial to see why this definition is a good model of the CHC problem, part of the reasoning is that using the definition of $A(n) = n < 100$, the query CHC \autoref{eq:me_q1} takes the form $\bot \leftarrow n < 100 \land n \geq 100$. The body of this CHC is clearly unsatisfiable, thus, $\bot$ cannot be deduced.
\end{example}

\subsubsection{Refutation Creation}
\label{sec:forward_refut}

When a CHC problem is unsatisfiable, a deduction can be found from the facts to $\bot$ that is always valid, regardless of how the uninterpreted functions are defined. The refutation is then a series of applications of the CHCs in the CHC set that start with a fact CHC and end with a satisfiable query CHC.

The counterexample provided by the model checker is an alternating sequence of concrete states of the CFA and edges. It starts at the initial location of the CFA with some values assigned to the variables and ends in the error location. The transformation described in \autoref{sec:forward_t} ensures that a location $l_i$ in the CFA can only be reached with the related variables $b_1^i, b_2^i, \dots, b_{m_i}^i$ valued $x_1, x_2, \dots, x_{m_i}$ if, and only if $B_i(x_1, x_2, \dots, x_{m_i})$ can be deduced. Consequently, all predicates corresponding to the locations of the concrete states in the counterexample are deducible, with the valuations present in the concrete states as parameters. The transformation also creates a one-to-one mapping of CHCs and edges. Thus, mapping the edges in the counterexample back to their CHCs, with the values of variables in the concrete states substituted as parameters, amounts to a valid refutation of the CHC problem's satisfiability.

\begin{example}
Since the motivating \autoref{ex:mot} is satisfiable, consider a modified version of it, in which the only fact is replaced with $A(n) \leftarrow n > 0 \land n \leq 100$. The forward generated CFA would be similar to the one in \autoref{fig:forward_ex}, with the exception of the edge going from $l_{Init}$ to $l_A$ having $n \leq 100$ instead of $n < 100$ in its guard.

The model checking algorithm would return the following counterexample, with the irrelevant variable values omitted:
\begin{align*}
    &(l_{Init}, n = 100) \\
    &(l_{Init}, ([n > 0, n \leq 100], a_1 = n), l_A) \\
    &(l_A, n = 100, a_1 = 100) \\
    &(l_A, (n = a_1, [n >= 100]), l_{Err}) \\
    &(l_{Err}, n = 100, a_1 = 100)
\end{align*}
This could be mapped to the refutation below:
\begin{align*}
    A(n) &\leftarrow (n > 0 \land n <= 100) \land n = 100 \\
    \bot &\leftarrow (A(n) \land n \geq 100) \land n = 100
\end{align*}
Since all variables have values assigned to them, it is trivial to check that this is indeed unsatisfiable.
\end{example}

\section{Evaluation}
\label{sec:evaluation}
\paragraph{Implementation} The CHC to CFA transformation steps were implemented as ANTLR frontends~\cite{antlr} in the open-source model checking framework \textsc{Theta}~\cite{theta}. The implementation is able to check the satisfiability of a CHC problem; however the generation of refutations and proofs is not implemented yet. 
Backward transformation was also implemented in a similar manner in the tool for comparison.
\begin{table}[b]
	\centering \scalebox{0.8}{
    	\begin{tabular}{ | c c c | c c | }
    	    \hline
    		\multirow{2}{*}{domain} & \multirow{2}{*}{interpolation} & \multirow{2}{*}{pred-split} & \multicolumn{2}{c | }{transformation} \\
    		 & & & BACKWARD & FORWARD \\
    		\hline
    		EXPL        & NWT\_IT\_WP  & -     & 77  & 138 \\
    		EXPL        & NWT\_WP\_LV  & -     & 82  & 137 \\
    		EXPL        & SEQ\_ITP     & -     & 81  & 175 \\
    		PRED\_BOOL  & BW\_BIN\_ITP & WHOLE & 110 & 288 \\
    		PRED\_CART  & BW\_BIN\_ITP & WHOLE & 141 & 302 \\
    		PRED\_SPLIT & SEQ\_ITP     & ATOMS & 131 & 310 \\
    		PRED\_SPLIT & SEQ\_ITP     & WHOLE & 142 & 318 \\
    		PRED\_SPLIT & BW\_BIN\_ITP & ATOMS & 83  & 291 \\
    		PRED\_SPLIT & BW\_BIN\_ITP & WHOLE & 114 & 328 \\
    		\hline
    	\end{tabular}
        }
	\caption{Number of solved tasks by certain configurations.}
	\label{table:z3configs}
\end{table}

\paragraph{Goals and Design} The aim of this evaluation is to show the effectiveness of the bottom-up approach by comparing it to the top-down approach. It also aims to study the performance of the approach with different  configurations of CEGAR, e.g.,\ different abstract domains.

The main comparison was done inbetween configurations of \textsc{Theta} only. Thus we were able to compare the different transformation approaches while the verification process was the same. Additionally, we also compared \textsc{Theta} to other state-of-the-art CHC solvers.

The implementation was evaluated on 585 linear CHCs over the background theory of linear integer arithmetic from the LIA-Lin track of the CHC-COMP21 benchmark repository\footnote{\href{https://github.com/chc-comp/chc-comp21-benchmarks}{https://github.com/chc-comp/chc-comp21-benchmarks}}.  The benchmarks were run on machines with 8 logical CPU cores and 16 GB of memory, with a timeout of 300 seconds.

\paragraph{Results}

\autoref{table:z3configs} shows the results of \textsc{Theta} with the different configuration options of \textsc{Theta}~\cite{theta}. The results of the tool were either correct or timeout for all of the tasks.

Forward transformation proved to be far more effective than backward transformation in all configurations. The configurations using boolean predicate based abstraction with sub-state splitting (\texttt{PRED\_SPLIT}) performed the best, with the other predicate based abstraction methods not too far behind.

The same benchmarks were also run with the top solvers of the LIA-Lin track from CHC-COMP21 \cite{chccomp21}, namely \textsc{Z3}, \textsc{Unihorn} and \textsc{Eldarica}. These solvers were run using their default configuration and with the same constraints as \textsc{Theta}. \autoref{table:othertools} shows the number of solved tasks compared to the best-performing configuration of \textsc{Theta}. Although \textsc{Theta} performs worse than the other solvers, its performance is comparable to \textsc{Eldarica}'s.

\begin{figure}
\begin{minipage}[c]{0.25\textwidth}
    \begin{table}[H]
	\centering
        \scalebox{0.8}{
    	\begin{tabular}{ | r | l | }
    	    \hline 
                \textbf{\textsc{Theta} (FW)} & \textbf{328} \\
    		\textsc{Theta} (BW) & 142 \\
                \textsc{Eldarica} & 337 \\
                \textsc{Unihorn} & 380 \\
                \textsc{Z3} & 437 \\
                \hline
    	\end{tabular}
        }
	\caption{Comparison to other tools.}
	\label{table:othertools}
\end{table}
\end{minipage}%
\begin{minipage}[c]{0.75\textwidth}
\begin{figure}[H]
	\centering
	\pgfplotstableread[col sep=semicolon,read comma as period=true]{figures/chc-quantile-data.csv}\datatable
\begin{tikzpicture}[scale=1, every node/.style={transform shape}]

\tikzset{mymark/.style={
        decoration={
            markings,
            mark= between positions 0 and 1 step 5mm with
                {
                \pgfuseplotmark{#1};
            },
        },
        postaction={decorate}
    }
}

\begin{axis}[
width=10cm,
height=7cm,
xtick distance={50},
x tick label style={font=\normalsize, rotate=0, anchor=north},
legend style={at={(0.97,0.05)},anchor=south east,font=\footnotesize},
ylabel={Time (s)},
ymode=log,
log ticks with fixed point,
grid=both,
grid style={gray!20},
xmin=0,xmax=450,
ymin=0.01,ymax=1000,
every axis plot/.append style={ultra thick}]

\addplot [red!80, mark=*, mark repeat=40, /pgf/number format/read comma as period=true ] table [x expr=\coordindex, y={Theta (FW)}]{\datatable};
\addlegendentry{\textsc{Theta} (FW)}

\addplot [orange!80, dashed, /pgf/number format/read comma as period=true ] table [x expr=\coordindex, y={Theta (BW)}]{\datatable};
\addlegendentry{\textsc{Theta} (BW)}

\addplot [green!80!black, mark=square*, mark repeat=40, /pgf/number format/read comma as period=true ] table [x expr=\coordindex, y={Eldarica}]{\datatable};
\addlegendentry{\textsc{Eldarica}}

\addplot [black!80, mark=diamond*, mark repeat=40, /pgf/number format/read comma as period=true ] table [x expr=\coordindex, y={Unihorn}]{\datatable};
\addlegendentry{\textsc{Unihorn}}

\addplot [blue!80, mark=triangle*, mark repeat=40, /pgf/number format/read comma as period=true ] table [x expr=\coordindex, y={z3}]{\datatable};
\addlegendentry{\textsc{Z3}}

\addplot [red!80, mark=*, mark repeat=40, /pgf/number format/read comma as period=true ] table [x expr=\coordindex, y={Theta (FW)}]{\datatable};

\end{axis}
\end{tikzpicture}
	\caption{Number of solved tasks by tools under a certain time.}
	\label{fig:br_others_q}
\end{figure}
\end{minipage}%
\end{figure}

A quantile plot of the tools' performances can be seen on \autoref{fig:br_others_q}. \textsc{Theta} performs better than both \textsc{Unihorn} and \textsc{Eldarica} for easier tasks, but it starts to get slower at a faster pace for tougher tasks than the other tools.

\paragraph{Conclusion} As shown in \autoref{table:z3configs}, the performance of \textsc{Theta} was greatly improved by the forward transformation process for all of the tested configurations. This improvement gains even more significance when compared to other tools: just by changing the transformation method, \textsc{Theta} becomes a relevant competitor for some of the best linear CHC solvers of CHC-COMP. Based on our findings, we propose that tools employing software verification techniques for CHC solving implement our novel approach, to potentially significantly increase the number of successfully solved CHC problems.

\begin{footnotesize}
\paragraph{\footnotesize{Acknowledgement.}}
We want to thank Martin Blicha for the help and fruitful discussion regarding the topic.
\paragraph{\footnotesize{Funding.}}
This research was partially funded by the \'UNKP-22-\{2,3\}-I New National Excellence Program and the 2019-1.3.1-KK-2019-00004 project from the National Research, Development and Innovation Fund of Hungary.
\end{footnotesize}
\bibliographystyle{eptcs}
\bibliography{generic}
\end{document}